\documentclass{PoS}
\usepackage{graphicx}
\usepackage{wrapfig}

\title{Detection of High Energy Cosmic Rays at the Auger Engineering Radio Array}

\ShortTitle{Auger Radio Detection}

\author{\speaker{Sijbrand de Jong} \hspace{1mm} for the Pierre Auger Collaboration
        \thanks{Observatorio Pierre Auger, Av. San Martín Norte 304, 5613, Malarg\"{u}e, Argentina}
        \hspace{0.1mm}
        \thanks{Full author list: \texttt{http://www.auger.org/archive/authors\_2016\_08.html}}
        \\
        Radboud University Nijmegen and Nikhef\\
        E-mail: \email{sijbrand@hef.ru.nl}
}

\abstract{
Detection of (ultra-) high-energy cosmic rays with the use of radio
frequency emission from extensive air showers has been proven as
complimentary to existing ground array detection techniques.
Great progress has been made in the understanding of
the emission processes and in their Monte Carlo modelling. These
have led to experimental results, notably also at the Auger
Engineering Radio Array, on energy, angular and mass composition
resolution of the primary cosmic rays. Recent results are reported.
The measured resolutions turn out to be competitive with existing
techniques and they have independent uncertainties. Implications
for cross-calibration with existing techniques and for future
large-scale applications of radio detection of ultra-high-energy cosmic
rays are discussed.
}

\FullConference{38th International Conference on High Energy Physics\\
		3-10 August 2016\\
		Chicago, USA}

\begin{document}
\vspace*{-17mm}
\section{Ultra-High-Energy Cosmic Rays and the Pierre Auger Observatory}
\vspace*{-3mm}
The Pierre Auger Observatory~\cite{AugerObservatory} consists of a 3000~km$^2$
area near Malarg\"{u}e, Argentina,
covered by more than 1660 water-Cherenkov stations in an hexagonal grid
with 1.5~km spacing. The area is surrounded by four telescope
stations, with 6 fluorescence telescopes each. Near one of the telescope stations,
there is
an area with several enhancements, including a higher water-Cherenkov station density,
underground muon detectors and a radio detector array.
The latest results on ultra-high-energy cosmic rays from
the Telescope Array Collaboration~\cite{TAtalk} and
the Pierre Auger Observatory~\cite{Augertalk} have been presented at this conference.
In this contribution, the new technique of radio detection of cosmic rays is discussed,
mostly based on the results of the Auger Engineering Radio Array (AERA)~\cite{AugerObservatory}.
The basic quantities that can be measured of a cosmic ray are its energy,
arrival direction, and particle type. The latter is also referred to as composition
or mass of the cosmic ray.

\vspace*{-4mm}
\section{Radio Detection of Cosmic Rays}
\vspace*{-3mm}
When an extensive air shower develops it emits electromagnetic radiation,
with an important part of the spectrum in the 10-100~MHz range. For short this
is called {\em radio emission}.
There are two contributing emission mechanisms:
geomagnetic emission due to the deflection of the electrons and positrons when traversing
the Earth magnetic field,
and emission due to charge separation in the shower that is caused by the ionisation electrons that
tend to move with the shower front, leaving behind the positively
charged ions and the annihilation of the positrons in the shower. The geomag-
\\[-5.5mm]
\begin{wrapfigure}{r}{0.4\textwidth}
\vspace*{-6mm}
\centerline{\includegraphics[width=0.335\textwidth]{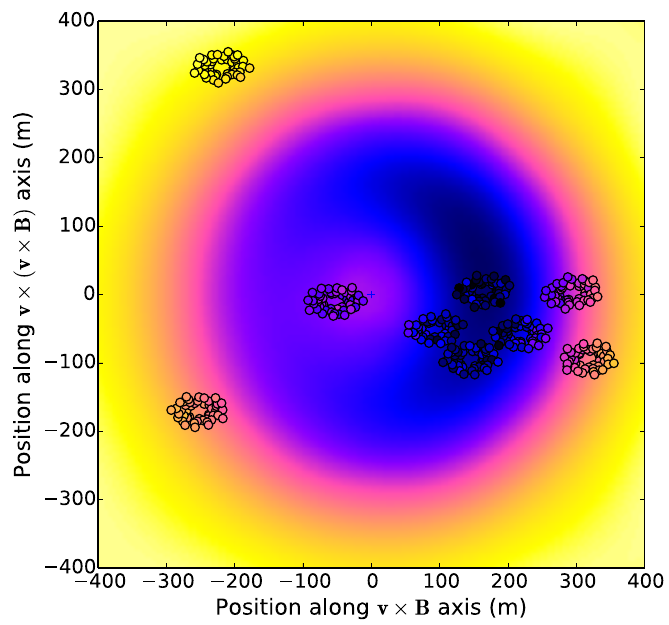}}
\vspace*{-3mm}
\caption[]{\label{LOFARXmaxFit}
               Lateral signal density distribution in the
               $\vec{v}\times\vec{B}$-$\vec{v}\times\vec{v}\times\vec{B}$
               plane around the shower axis position at $(0,0)$ in
               the LOFAR observatory~\cite{LOFARXmax}.
               The coloured circles are the data and the coloured
               background represents a heuristic model.
               The fit is perfect if the colours fully blend.
               }
\end{wrapfigure}
netic emission has 
the same polarisation at all  points in space, perpendicular to the
Earth magnetic field ($\vec{B}$) and the shower axis ($\vec{v}$),
the $\vec{v}\times\vec{B}$ direction.
The charge separation emission is polarised radially towards the shower axis.
The emission occurs in a long path along the shower axis. This causes additional interference
between early and late emission due to the finite refractivity of air for radio waves,
resulting in a Cherenkov cone structure of the signal around the show axis.
The two emission contributions interfere by vector addition in any given point in space,
resulting in an asymmetric signal pattern around the shower axis.
Both the Cherenkov ring in the signal amplitude, as well as the asymmetry around the shower
axis are illustrated by a measurement presented in Fig.~\ref{LOFARXmaxFit}.

\vspace*{-4mm}
\section{The Auger Engineering Radio Array}
\vspace*{-3mm}
\begin{figure}[htbp]
\vspace*{-5mm}
\hfill
\includegraphics[height=5cm]{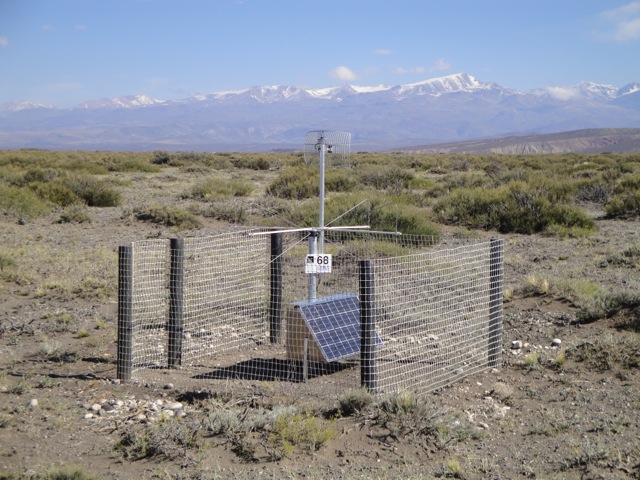}
\hfill\hfill
\includegraphics[height=5cm]{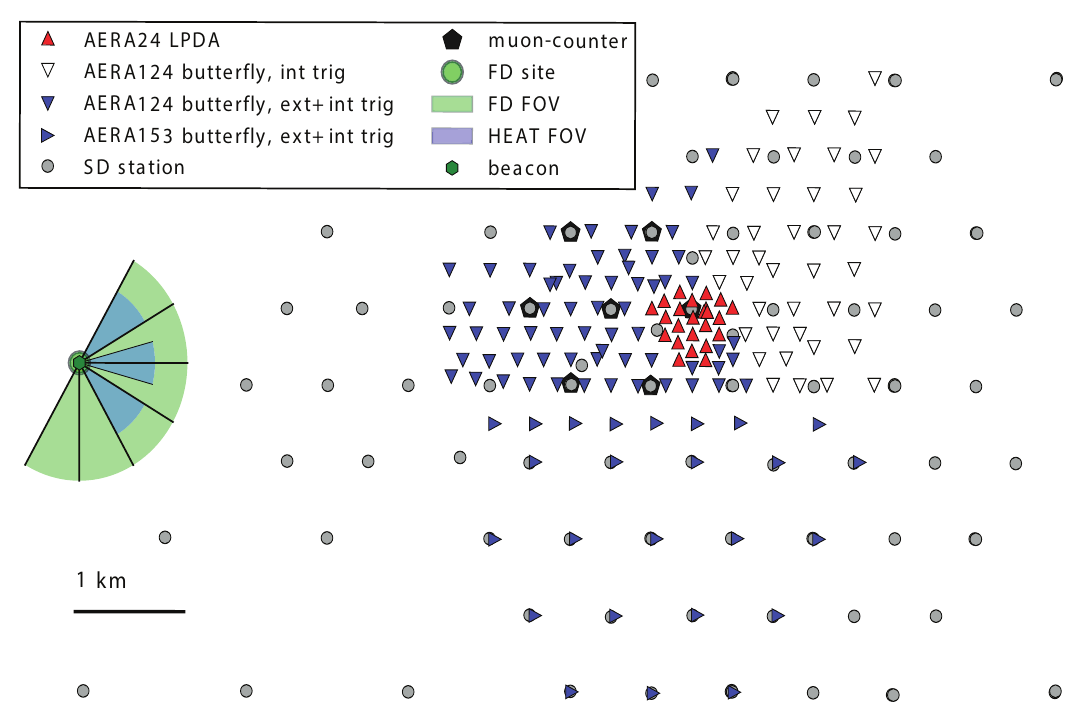}
\hfill
\vspace*{-3mm}
\caption[]{\label{AERAlayout}
              Left: A butterfly radio detector station in the field. Right: Position of the radio detector
              stations (triangles) in relation to the fluorescence detector on the left and the surface
              detector stations drawn with grey circles. The fluorescence building also sites a beacon
              to time-calibrate the radio stations.
              }
\vspace*{-3mm}
\end{figure}
The Auger Engineering Radio Array consists of dipole antennas with the appropriate length
 to receive
30-80 MHz radiation. The left hand side of Fig.~\ref{AERAlayout} shows the butterfly design,
which applies to the majority of the stations in the array.
Each detector station has digitizing electronics, is autonomously powered by a solar panel and
battery, and communicates wirelessly. A GPS receiver takes care of a time stamp with an event.

The array was built in three phases: first 24 Log-Periodic Dipole Antennas (LPDA) were set up
at relative small interstation spacing, shown as the red triangles in Fig.~\ref{AERAlayout}
on the right.
In a second phase, butterfly antennas were added with two different electronics designs,
one aimed at external triggering by the Auger surface and fluorescence detectors 
(blue triangles pointing down in Fig.~\ref{AERAlayout}) and
one aimed primarily at self-triggering (open triangles in Fig.~\ref{AERAlayout}).
In a third phase antennas were added recently at larger interstation spaces 
(blue triangles pointing right in Fig.~\ref{AERAlayout}), bringing the total
number of antennas deployed in the array to 153 detector stations.

The signal voltage $U$ in a 30-80 MHz frequency window is determined from the digital time trace
using a Hilbert envelope. Within the frequency window some frequency bands with anthropogenic
noise are suppressed.
The measured signal voltage is related to the electric field $\vec{E}$ of the radio emission
by $U_{{\rm pol}\,1,2}=\vec{\vec{H}}_{{\rm pol}\,1,2} \vec{E}$, where $\vec{\vec{H}}_{{\rm pol}\,1,2}$
is the antenna pattern for the polarisation directions 1 and 2.
The antenna pattern can be calibrated by antenna simulation,
by calibration with a point source, and by using the galactic center that emits radio frequency signals.
All these calibration methods are being used in AERA.
The energy density $\epsilon$ in eV/m$^2$ is then measured by integrating the electric field squared
over a signal window from $t_1$ to $t_2$
and subtracting a similar integral over a signal-free background time window preceding
the signal from $t_3$ to $t_4$:
$\epsilon=\epsilon_0 c \left[ \sum_{t_i=t_1}^{t_2} |\vec{E}(t_i)|^2 \Delta t
                                           - \frac{t_2-t_1}{t_4-t_3}\sum_{t_i=t_1}^{t_2} |\vec{E}(t_i)|^2 \Delta t
                                    \right]$,
where $\epsilon_0$ is the dielectric constant in vacuum, $c$ the speed of light and
$\Delta t$ the width of the time bins.

\vspace*{-3mm}
\section{Results}
\vspace*{-3mm}
\subsection{Emission Mechanisms}
\vspace*{-1mm}
The good agreement of the predictions of the signal distribution compared to data
such as presented in Fig.~\ref{LOFARXmaxFit}
shows that the emission mechanisms, their interference, and the propagation through the
atmosphere of the radiation are well understood.

The two emission mechanisms may be disentangled using the polarization pattern in the
$\vec{v}\times\vec{B}$-$\vec{v}\times\vec{v}\times\vec{B}$ plane.
The total electric field is given by
$\vec{E}=\vec{E}^{\rm G}+\vec{E}^{\rm C}=|E^{\rm G}|\vec{e}^{\rm \;G}+|E^{\rm C}|\vec{e}^{\rm \;C}$,
with $^{\rm G}$ and $^{\rm C}$
denoting the geomagnetic and charge excess components, respectively.
The geomagnetic emission is proportional to the angle $\alpha$ of the shower axis with
the geomagnetic field $|E^{\rm G}| \propto |\sin\alpha|$.
The average charge excess signal relative to the geomagnetic signal is then found as
$a=\sin\alpha \frac{|E^{\rm C}|}{|E^{\rm G}|}$.
For a certain selection of AERA events~\cite{AugerRadioPol}, $a=14\%$,
a small but non-negligible contribution.

\vspace*{-2mm}
\subsection{Energy Measurement}
\vspace*{-1mm}
For the energy measurement the signal is fit by an heuristic parametrisation
that reproduces the the lateral distribution well
$\epsilon(\vec{r})=A\left[
{\rm e}^{-(\vec{r}+C_1\vec{e}_{\vec{v}\times\vec{B}}-\vec{r}_{\rm core})^2/\sigma^2}
-C_0
{\rm e}^{-(\vec{r}+C_2\vec{e}_{\vec{v}\times\vec{B}}-\vec{r}_{\rm core})^2/(C_3 {\rm e}^{C_4 \sigma})^2}
\right]$.
\begin{wrapfigure}{r}{0.46\textwidth}
\vspace*{-7mm}
\centerline{\includegraphics[width=0.4\textwidth]{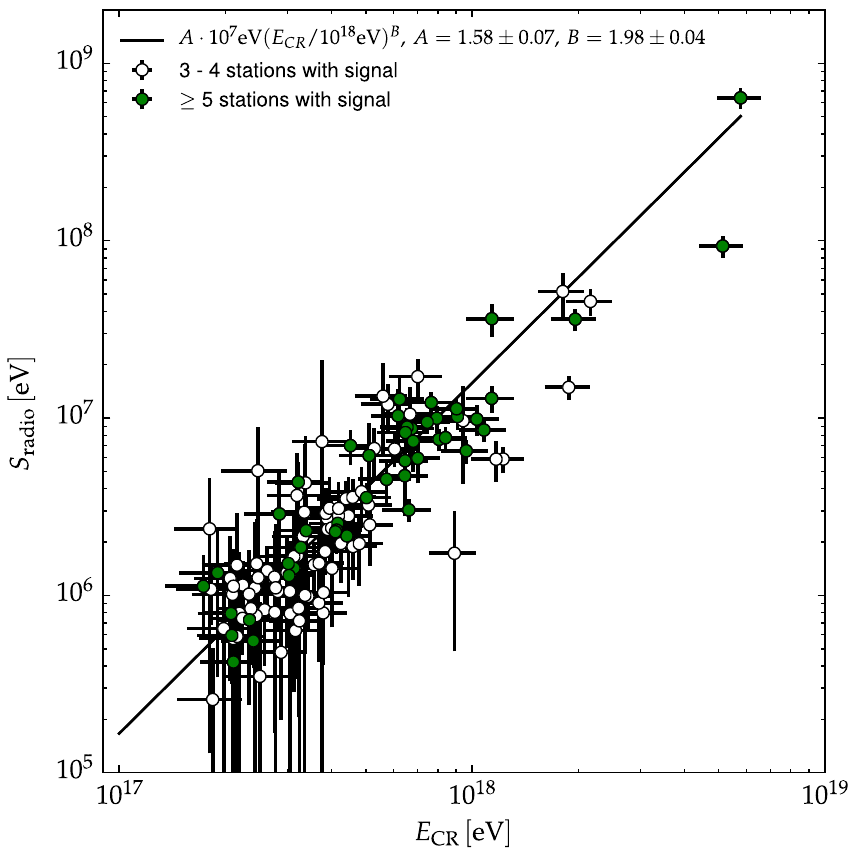}}
\vspace*{-3mm}
\caption[]{\label{RDFDenergy}
               Total energy measured in the emission in the 30-80~MHz frequency band on
               the vertical axis compared to the energy measured for the same shower by the
               fluorescence detector~\cite{AERAEnergy}.
               }
\vspace*{7mm}
\end{wrapfigure}

\vspace{-12mm}
\noindent
The parameters $C_i$ are obtained from Monte Carlo simulation and are fixed for all data events.
The core
position of the shower, $\vec{r}_{\rm core}$, the total signal strength $A$, and a measure of
the width of the shower, $\sigma$, are fit for each shower.
See Fig.~\ref{LOFARXmaxFit} for such 
a fit on an event measured in the LOFAR observatory.
The total energy in the radio frequency emission in the 30-80~MHz range for each events is
then obtained as
$S_{30{\rm -}80\,{\rm MHz}}=\frac{1}{\sin^2\alpha}\int\epsilon(\vec{r}){\rm d}\vec{r}=
\frac{\pi A}{\sin^2\alpha} \left[\sigma^2-C_0C_3^2{\rm e}^{2C_4\sigma}\right]$.
To first order the contributions of the charge excess emission cancel out in the integral and the
total signal strength is determined by the geomagnetic emission.

The comparison of the total energy measured in the radio signal to the energy observed
by the fluorescence detectors of the Pierre Auger Observatory is shown in Fig.~\ref{RDFDenergy}.
From the slope in this figure it is measured that the total measured radio frequency emission energy
as a function of the cosmic-ray energy, $E_{\rm CR}$ and the Earth magnetic field strength, $B$, is
given by
$S_{30{\rm -}80\,{\rm MHz}}=(15.8\pm 0.7\pm 6.7)\,{\rm MeV}\;
\left(\sin\alpha\;\frac{E_{\rm CR}}{10^{18}\;{\rm eV}}\;\frac{B}{0.24\;{\rm G}}\right)^2$, where
the dominant systematic uncertainties are in nearly equal amount due to the uncertainty in the
unfolding of the antenna pattern and energy scale uncertainty of the fluorescence
detector~\cite{AERAEnergy}.
The shower-to-shower uncertainty in the energy measurement from the radio signal is estimated to
be 22\% when at least three radio detector station are contributing to the measurement, going
down to 17\% for showers with five or more detector stations contributing to the fit~\cite{AERAEnergy}.

Work is progressing to relate the energy of the extensive air shower to the energy in 
radio emission. This potentially allows an absolute calibration of cosmic ray energy
that is completely independent from the fluorescence detectors and even may allow
to fix the absolute energy scale of the fluorescence detector to a much better precision
than currently known.

\vspace*{-3mm}
\subsection{Direction Measurement}
\vspace*{-2mm}
The arrival direction of the cosmic ray is determined by the direction perpendicular to the shower plane,
which requires excellent relative detector station timing.
This relative timing is calibrated and tracked using the signal from a beacon that is installed
on the nearby fluorescence telescope building. The beacon emits at four fixed
frequencies, causing a beat signal with period 1100 ns.
The relative station time is calibrated 
 to the  
\\[-5mm]
\begin{wrapfigure}{r}{0.4\textwidth}
\vspace*{-11mm}
\centerline{\includegraphics[width=0.4\textwidth]{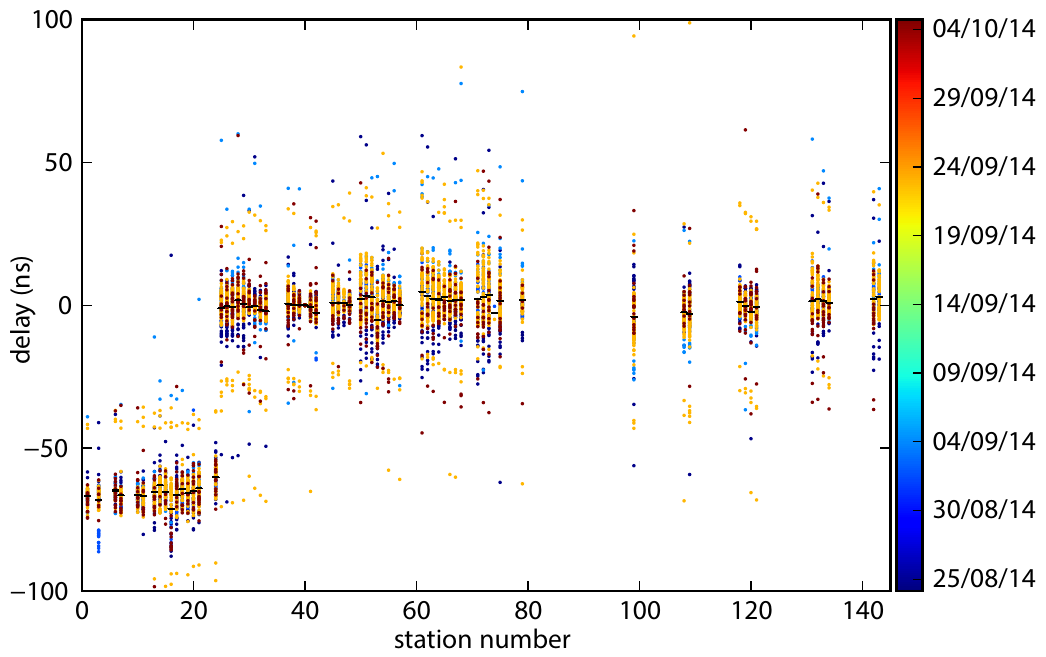}}
\vspace*{-3mm}
\caption[]{\label{PlaneTiming}
               Timing offsets for signals from airplanes as a function of detector station
               number~\cite{AERAAirplane}.
               }
\vspace*{-4mm}
\end{wrapfigure}
phase of this beat signal with respect to one chosen
station~\cite{AERAAirplane}.
Some air planes that emit signals in the 30-80 MHz range were used as point sources,
because their position in time can be precisely followed.
From these signal pulses, the relative timing of the stations was checked, as illustrated in
Fig.~\ref{PlaneTiming}. This figure shows that the timing resolution is of the order
of a nanosecond for individual measurements.
This procedure also revealed a fixed time offset between the LPDA and butterfly antennas
of which the origin is not yet understood~\cite{AERAAirplane}.
With a timing resolution of 1~ns an angular resolution for the cosmic ray of about $0.1^{\circ}$ is
theoretically possible. The airplanes were tracked with a precision better than $1^{\circ}$.

\vspace*{-2mm}
\subsection{Particle Type Measurement}
\vspace*{-2mm}
The primary observable to discern the different particle types of cosmic rays is the
penetration depth in the atmosphere or position where the shower contains the maximum 
number of particles, $X_{\rm max}$, which is measured in g/cm$^2$.
So far only $X_{\rm max}$ measurements from the fluorescence detector have been published.
For the radio signal, several features of the signal pattern on the ground are sensitive to $X_{\rm max}$
or the nucleus type of the cosmic ray.
Using the lateral energy density profile, the width of the shower is the feature that is
most sensitive to $X_{\rm max}$.
In LOFAR, a method using the quality of the fit of the Monte Carlo simulations to the data
for various $X_{\rm max}$ trails has been developed~\cite{LOFARXmax}.
\begin{figure}[tbp]
\vspace*{-5mm}
\hfill
\includegraphics[height=4.5cm]{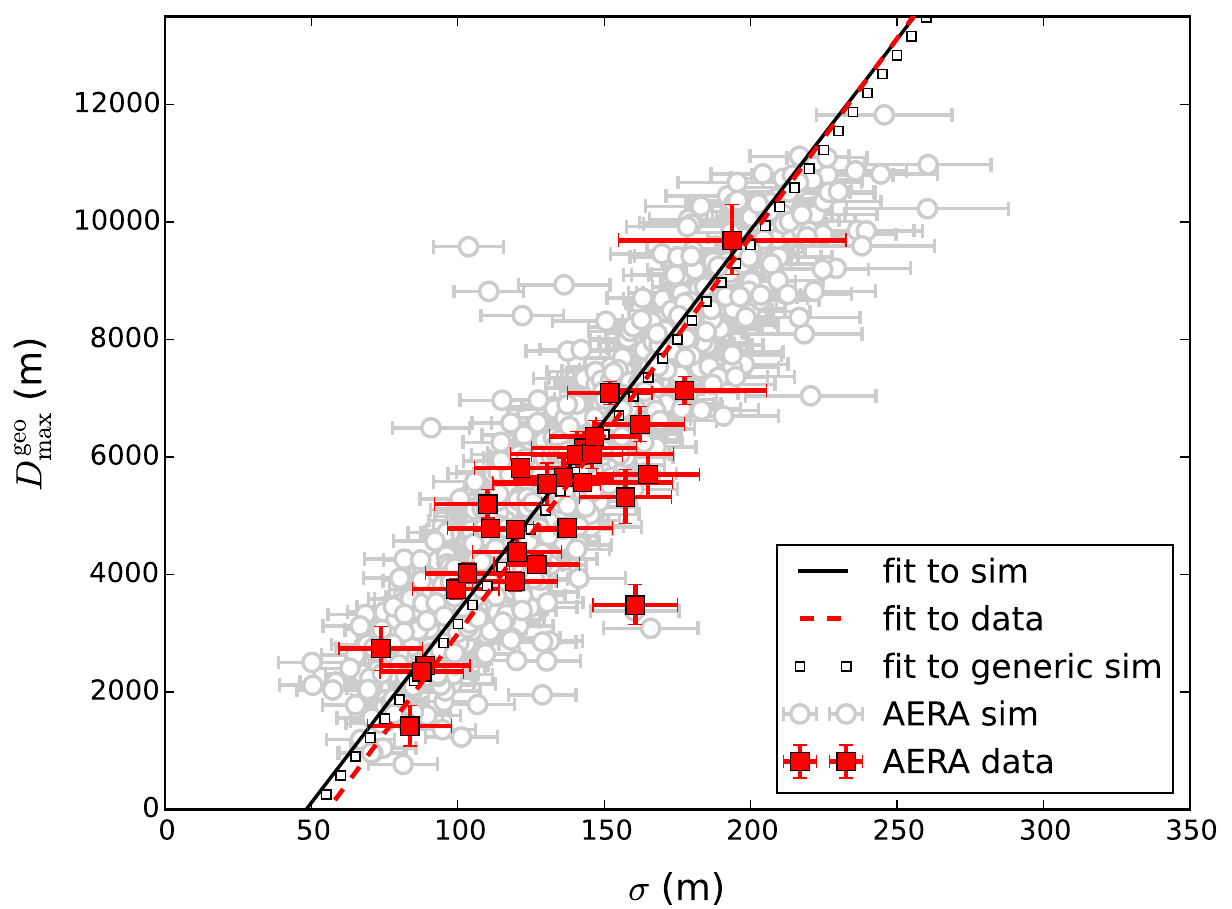}
\hfill\hfill
\includegraphics[height=4.5cm]{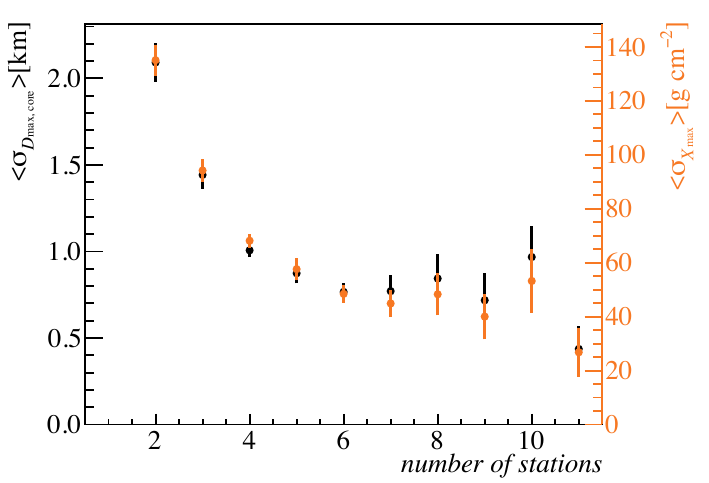}
\hfill
\vspace*{-3mm}
\caption[]{\label{RadioXmax}
              Left: $X_{\rm max}$ correlation found with the method of~\cite{LOFARXmax}
              for AERA data~\cite{SchulzThesis}.
              Right: Uncertainty on the $X_{\rm max}$ determination using the
              spectral index method~\cite{JansenThesis}.
              }
\vspace*{-2mm}
\end{figure}
This method has also been applied to AERA data in~\cite{SchulzThesis}, where a comparison
was made to the $X_{\rm max}$ that was found for the same showers using the fluorescence
detector. In Fig.~\ref{RadioXmax} this comparison is shown. It results in a resolution of
about 42~g/cm$^2$.

Another method to determine the $X_{\rm max}$ from radio signals uses the spectral index
from each individual detector station~\cite{JansenThesis}. The results from the individual
stations are then combined into a result for the shower.
On the right in Fig.~\ref{RadioXmax} the uncertainty of this spectral index
method is shown as a function of the number of active detector stations, and can get
as good 40~g/cm$^2$.

A third method that uses the arrival time at each radio detector to accurately fit the curved
wave front, thereby allowing one to infer the position of $X_{\rm max}$ is under study in AERA.
The three methods use independent information and can be compared to study systematic
uncertainties and combined to deliver better accuracy.

\vspace*{-2mm}
\subsection{Horizontal Showers}
\vspace*{-2mm}
\begin{wrapfigure}{r}{0.44\textwidth}
\vspace*{-14mm}
\centerline{\includegraphics[width=0.35\textwidth]{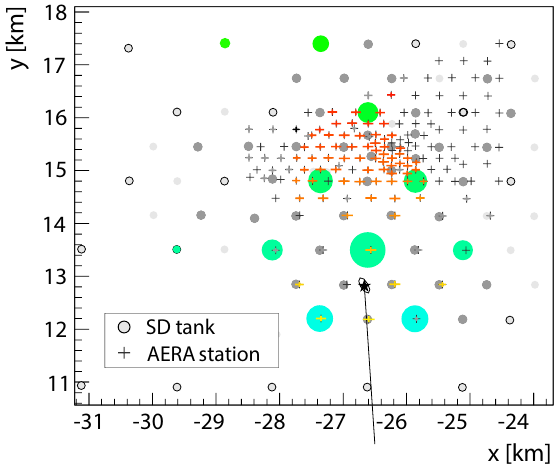}}
\vspace*{-3mm}
\caption[]{\label{HorizontalShower}
               Shower at 74.4$^{\circ}$ zenith angle hitting 69 AERA stations indicated by
               the coloured + signs~\cite{ARENAHorShowers}.
               }
\vspace*{-4mm}
\end{wrapfigure}
Radio detectors are also sensitive to horizontal showers, which make them also well suited for
ultra-high-energy neutrino detection. As a proof of principle, horizontal showers in  AERA have
been shown in~\cite{ARENAHorShowers}. Figure~\ref{HorizontalShower} shows such an event
with a zenith angle of 74.4$^{\circ}$ that produced  a very elongated footprint,
hitting 69 radio detector stations. Such showers can also be well
reproduced with CoREAS simulations~\cite{ARENAHorShowers,CoREAS}.

\newpage
\vspace*{-18mm}
\section{Conclusion and Future Developments}
\vspace*{-4mm}
Radio detection of cosmic rays has been established as a viable technique
to measure the energy, direction, and particle type of (ultra-) high-energy cosmic rays.
The signal emission has been understood and is well modelled in several Monte Carlo
codes~\cite{CoREAS,ZHAires,EVA}.
The one fundamental property that still has to be understood is the efficiency of the
detection method.

For future large-scale applications, e.g.\ for ultra-high-energy neutrino
detection~\cite{GRAND}, in addition to establishing the detection efficiency,
more work is needed to make a robust trigger on the radio data alone.
Also major effort still has to be put into making the detector station more robust
and much cheaper. For the moment there are no apparent show stoppers to
achieve these goals.

\end{document}